# Laptop-assisted Helium Liquefier: software for tailored monitoring and control

Marco Marceddu

CGS - Centro Grandi Strumenti d'Ateneo, University of Cagliari, ITALY

marco.marceddu@dsf.unica.it

**Abstract**

Computer running human machine interfaces are fast supplanting conventional hardware dedicated to monitoring and supervising tasks. This kind of approach was successfully employed to develop a new monitoring and control software, running on a conventional laptop computer, for a small size helium liquefier. This software was realized with a SCADA development package and hallowed to manage all the liquefier functionalities. Being conceived as an open structure suitable for further upgrades, this liquefier HMI can be tailored on demand to satisfy specific needs of the end-user, even if not provided by the original manufacturer project.

**Introduction**

Helium recovery and liquefier plants may be a strategically relevant choice for all those institutions, such as universities and large research facilities, where liquid Helium (LHe) is routinely employed. Indeed, being a non renewable resource with an handful of sources all over the world, He needs a careful management, in particular in terms of usage and recovery [1-4].
Commercially available plants are typically controlled by sophisticated and dedicated hardware and software, that enable automatic operation and monitoring of the process. The main disadvantage is related to installation, maintenance and upgrading costs. Usually plant manufacturers provide closed-box hardware and control systems, with built-in human-machine interfaces (HMI) that perform pre-defined tasks. Typically the number and kind of available tools cannot be changed by the end-user, who has no chance to get hands in. Therefore when failure solving or customized solutions are needed, users are obliged to depend on the manufacturer assistance, with consequent time and money waste. In many situations a different approach may be preferred, since the operator wants to develop his own control tools, or solve problems independently, quickly and economically. Recently the substitution of hardware control panels with dedicated control software running on conventional computers, is becoming a quite common practice [5-7]. This approach has a long-standing tradition in universities and other research institutions where both educational and research experiments are organized by means of virtual instruments developed with the LabVIEW package [6, 8]. In the industrial field instead, the same idea is realized with new SCADA oriented packages. The term SCADA (Supervisory Control And Data Acquisition) generically refers to complex systems controlling data and commands flux exchange, between the several components of a field plant and the central master station [5-7]. Accordingly a SCADA software controls the communication between the central unit and the remote components of the plant, and allowed the operator to perform supervisory and control tasks [5-10]. SCADA is widely used in several industries, such as oil refineries, power plants or other manufacturing activities since, also due to the Internet revolution, remote control over long distances is nowadays a consolidated practice [5, 6, 9, 10]. Moreover SCADA is becoming popular also in research facilities, where a large number of measuring instruments and devices are interfaced in networks controlled by a central unit [11, 12]

In this work we present an HMI software realized to improve the control and monitoring capabilities of the He liquefier plant of University of Cagliari. This software was developed with a commercially available SCADA package and is a mimic of the original HMI, but with a more friendly and efficient organization, and with the possibility to implement specifically devised tools, not previously available.

**The Helium Liquefier**

The He liquefier plant operating at University of Cagliari is the model 1410 by PSI Incorporated. The system is equipped with a recovery and purify unit that enables also the employment of He evaporated from storage dewars. Liquid He production rate is at about 8l/hrs with pure He supply, and 4l/hrs with impure He supply; production rate can be doubled by means of liquid nitrogen pre-cooling. A brief discussion of the He liquefaction process in presented in the appendix.

The liquefier operation is supervised and controlled by a SCADA hardware system. The main unit is a programmable logic controller (PLC) model Koyo D4-450 by Automation Direct. This PLC has a total available memory of 30.8 K words, with contact execution time of 0.96 μs and typical scan time of 4-5 ms; 4 built-in communication ports allowed the connection with external devices by means of K-sequence (proprietary), DirectNET, Modbus or ASCII protocols, with a maximum baud rate of 38.4 K. The PLC is equipped with one coprocessor unit (Automation Direct F4-CP128-1) for process variable calculations. The SCADA network is also composed by auxiliary boards allowing the communication between the PLC and the remote terminal units (RTU) of the liquefier. One 8 channels analog input board with 12 bit AD converter (F4-08AD) manages the signals transmitted from gas pressure and temperature sensors and motor speed sensor; one 4 channels analog output board with 12 bit DA converter (F4-04DA) pilots the engine brake signal, the main pressure controller and the Joule-Thomson (JT) valve position; one 16 channels 110VAC input board (D4-16NA) receives status signals from the gas compressor, engine over speed sensor, gas supply sensor and fuses; two 16 channels 110-220 VAC output board (D4-16TA) are used to drive power absorbing RTU such as the compressor, motor brakes, valve actuators and the vacuum pump. The original HMI is a built touch-screen device (Cutler Hummer Panel Mate Power Series), communicating with the PLC, that enables the operator to call specific lines of the ladder program, thus activating specific functions and parameters monitoring. The block diagram of the liquefier SCADA system is shown in figure 1.

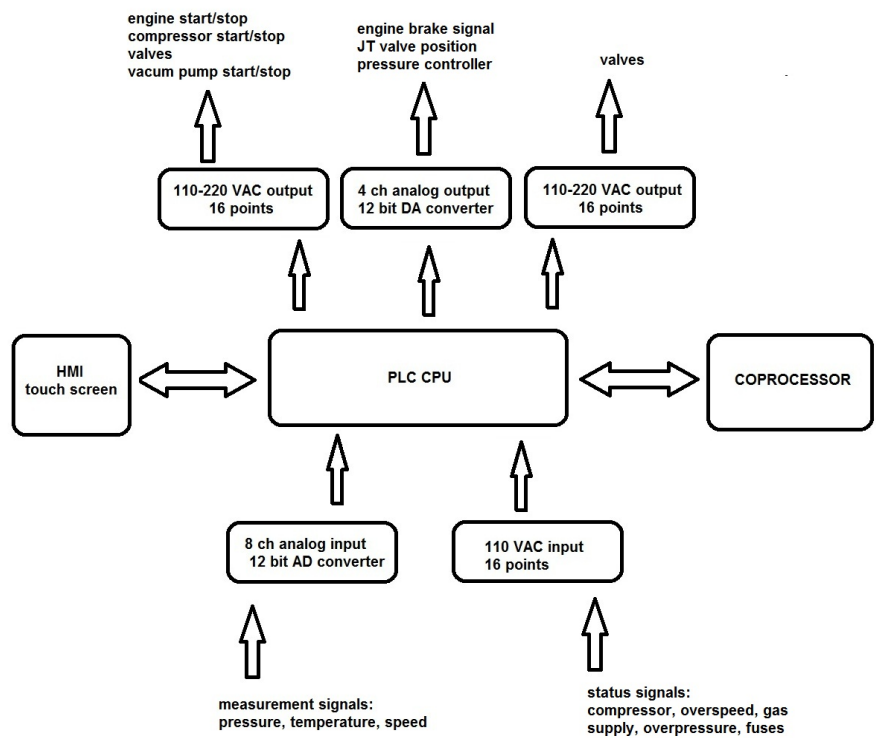

**Figure 1: block diagram of the liquefier SCADA system**

The PLC CPU is programmed by a relay ladder logic (RLL) type code. The main process variables are: engine speed, inlet and discharge gas pressure, gas pressure at JT valve, gas temperature at specific points of the flow diagram (close to the expansion engines and to the JT valve), JT valve opening level. Basically the main task of the PLC program is to keep the inlet gas pressure at the set point of 230 psi while keeping gas temperature at the first expansion engine at 19 K. This task is accomplished by a PID controller (proportional-integrative-derivative) that monitors and optimizes the process variables involved. The PLC program allows both manual and full automatic liquefier operation.

**The developed HMI software**

In projecting a new HMI system the ensuing requirements were taken into account: retaining all the original features of the primitive system; avoiding the employment of specialized and expensive hardware; networking capabilities; preference to industrial oriented solutions that offer long terms strength and reliability. Moreover this new HMI should be realized as an open structure, thus suitable for further improvements with new specifically devised tools. A functional and cost-saving solution was identified in interfacing the liquefier PLC with a laptop computer running a SCADA based user interface software. SCADA development packages are available both as open source or proprietary [5, 11]. Open source codes show appealing characteristics such as customising tools and multi-platform possibilities, but they may show not perfect matching with PLC protocols and may require advanced programming skill [11]. Therefore we preferred to retain a common platform with the original system, by using the SCADA package released by Automation Direct (POV - Point of View v7.1), in order to avoid compatibility and communication problems. Moreover POV language and architecture are very simple and intuitive, and don't require any specialized programming knowledge. Indeed POV is an object oriented package, where each object is associated with a variable (analogue, Boolean or string) called TAG, which is the core element of the POV language. TAGs may be input signals or bits, received by the PLC from the plant, internal bits or signals for local usage, output signals and bits driving plant element status. Available objects are indicators, controls, meters, buttons, switches, valve commands, sliders and many more. Supervising and monitoring functionalities are addressed by specific tools like data loggers, plots, reports, alarm and event notifications. A large variety of animation, graphical and acoustic tools are available to drawn the most useful and friendly user interface. A specific section of POV package is devoted to port and communication protocol definition. Indeed POV package is distributed with a library of communication drivers (DLL) suitable for several PLC. In particular it supports K-sequence protocol, which is the Automation Direct proprietary communication protocol. It is derived from DirectNET, thus being a serial RS232 based protocol.

The first step in writing the POV software was to compile a definition list of all the TAGs needed. In this case TAGs were identified with the process variables and bits employed by the liquefier PLC. The TAGs were then used to write specific expressions to program each object function.

The new HMI software was developed as a mimic of the original built-in HMI, but with significant differences addressed to a more efficient and friendly usage. Indeed being an engineered, market available system, the original HMI configuration was thought to meet just pre-defined, conventional needs. Thus if new, specifically tailored tools were needed, factory reconfiguration of the HMI was necessary. Moreover, since this kind of industrial liquefiers are devoted to very long, full-automatic runs, manual operation may result boring and tricky, with the operator obliged to navigate back and forth through different screens, to access specific data and tools. Therefore the new developed monitoring and control software was thought with a more rational format and with the possibility of easy and quick implantation of new tools, without the manufacturer assistance. It should be here specified that this new software can't be used to implement new routines of the liquefier process: these can be implemented only by acting on the PLC ladder logic code. Instead it enables a more friendly, reliable and customisable management of the process flow monitoring and control.

As the original HMI, the new software is organized in different screens, each one managing specific functions of the liquefier. In figure 2 is shown the program main screen (SCREEN 1). This screen shows all data and device status during liquefier running. In the top-left side panel of the screen are real-time displayed the engine speed, gas temperature (TEMP. A - TEMP. H) and pressure (PT1, PT2, PT60) values measured at different points of the He flow diagram, the percentage of JT vale opening. In the middle of the screen are shown the status indicators of the liquefier main components (compressor, expansion engines, valves and others), while in the bottom part of the screen are displayed alarm, event and error notifications (reset buttons are located on the left-side of this display). On the left side of the main screen seven buttons give access to all the other screens of the program, each one controlling specific liquefier components. In particular screen 7 gives access to the system parameter and PID configuration.

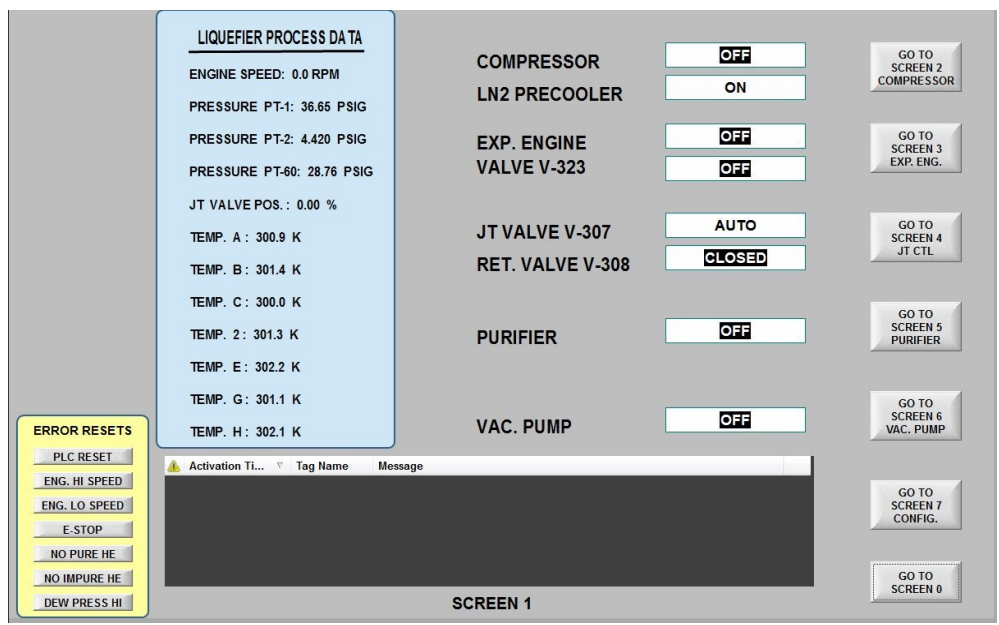

**Figure 2: HMI software main screen displaying process data and liquefier status**

As a further discussion of the software characteristics, figure 3 shows the expansion engine control screen. Push buttons are devoted to start and stop the engines and to switch between manual and automatic speed control. LED-like indicators advise the user about the engine status while the engine speed is shown as a process variable value in the speed controller panel. When manual speed control is activated, the speed set point is defined by the operator in the dedicated dialog box. Other indicators and controls of this screen pertain the inlet high pressure value and set point (PT1) and discharge low pressure value(PT2); specific panel is dedicated to He inlet main valve indicator and controls (valve V-323).

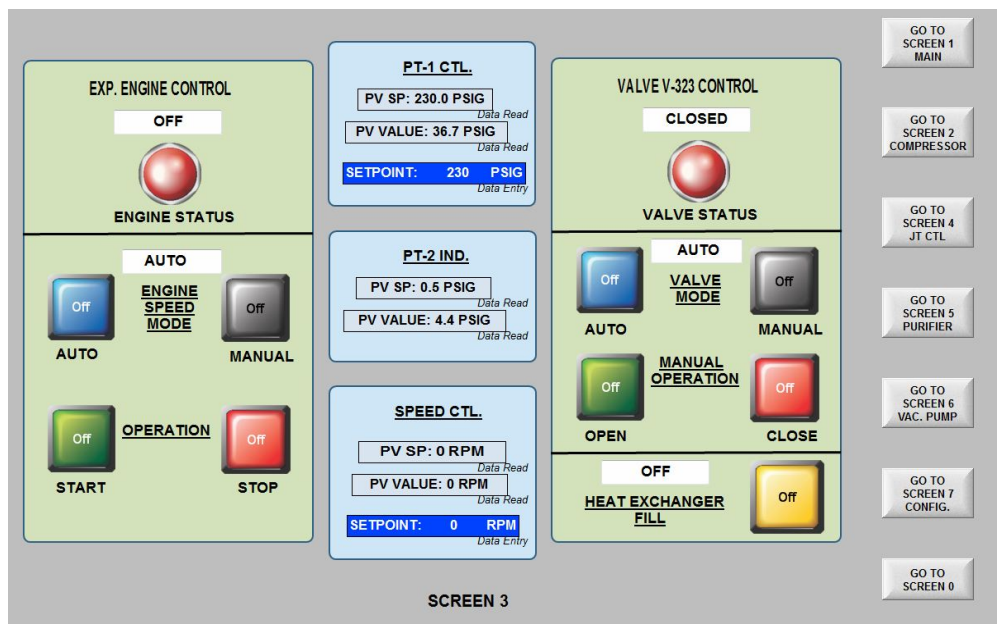

Figure 3: HMI software screen devoted to the expansion engine control

Debugging tests were carried out by comparing POV software performances with those displayed by the original built-in HMI, that was kept in communication with the PLC. These trials easily fixed minor problems, mainly due to TAGs misidentification or wrong number codification. Successively several liquefier runs (10-72 hours continuative working), were performed and no critical errors were ever verified. In particular the liquefier performances, both in term of process efficiency, rate, and supervision were perfectly preserved. Moreover the liquefier remote control via the Internet is available, improving the safety and efficiency of the process, while new tools, such as data logger and trend recorder, are in development.

In the last six month the POV HMI software has been exclusively and successfully used, as witnessed by the more than 600 litres of LHe produced, and at present is routinely used as the main liquefier control and supervisor system.

**Appendix: Helium liquefaction process**

PSI Incorporated model 1410 is a small-sized Helium liquefier, based on the Joule-Thomson effect, an isenthalpic expansion that, depending on the gas initial pressure and temperature, may result in gas temperature increase or decrease [1-3]. The initial condition corresponding to temperature decrease following throttling is called inversion temperature [1, 3]. In the model 1410 He cooling to the inversion temperature is realized by means of isentropic expansions, and of a counter flow cooling scheme. He flow diagram is composed of one inlet high pressure pipe and one return low pressure pipe. Along the high pressure line He experiences the two isentropic expansions and the isenthalpic expansion. The gas delivered from a screw compressor flows into the process line at the initial pressure of at about 1.6 MPa (230 psi); then it looses heat by means of work extraction during isentropic expansion, that takes place into two cylinder-piston assemblies (expansion engines). Exhaust gas then flows into the low pressure line where it is progressively warmed to room temperature by means of heat exchanging with further He flowing into the high pressure line, the latter one being pre-cooled before entering the expansion engines. This close loop continues until the gas temperature upstream the JT valve is below the He inversion temperature. Then He is allowed to flow trough the JT valve where isenthalpic expansion and further cool down to liquefaction temperature occurs, and liquid He is accumulated in a storage dewar. An auxiliary part of the liquefier is devoted to contaminated He purification. This purifier device is composed of cryogenics traps (cooled by the pure He main stream) where impurities condensate. This purified gas is then allowed to flow in the liquefier main line to be liquefied.

**Conclusions**

The He liquefier operating at University of Cagliari was recently upgraded by implementing a new supervision and control software, running on a conventional laptop computer. This new software perfectly matches all the functionalities of the original built-in terminal, but it is organized in a more friendly and rational format, thus allowing more practical and easy operation. In the same time the possibility to implement new specifically tailored control tools is now available.

The proposed approach shows how new object oriented programming packages can be used to meet specific needs and to develop new tools, even when these were not provided by the original project, and without calling for the manufacture assistance.

**Acknowledgements**

The author acknowledges Maksymilian Kozyrczak from Cryonova LLC, Tulsa, Oklahoma USA for the valuable technical support to this project.